\documentclass[preprint,12pt,authoryear]{elsarticle}

\usepackage{amssymb}
\usepackage{amsmath}
\usepackage{microtype}
\usepackage{lineno}
\usepackage{multirow}
\usepackage{booktabs}
\usepackage{url}
\usepackage{xurl}
\usepackage{hyperref}
\hypersetup{
  pdftitle={Uncertainty-Aware Crossmodal Fusion for Classification of Animal Behavior},
  pdfauthor={Ehsan Yaghoubi, Florian Haselbeck}
}
\usepackage{rotating}

\usepackage[colorinlistoftodos]{todonotes}

\journal{Computers and Electronics in Agriculture}

\begin{document}

\begin{frontmatter}



\title{Uncertainty-Aware Crossmodal Fusion for Classification of Animal Behavior} 


\author[label1]{Ehsan Yaghoubi\corref{cor1}}
\ead{ehsan.yaghoubi@hswt.de}

\author[label1]{Florian Haselbeck}
\ead{florian.haselbeck@hswt.de}

\cortext[cor1]{Corresponding author.}

\affiliation[label1]{organization={Weihenstephan-Triesdorf University of Applied Sciences},
            addressline={Department of Sustainable Agricultural and Energy Systems},
            city={Freising},
            street={Am Staudengarten 1,},
            postcode={85354},
            country={Germany}}

\begin{abstract}
Artificial intelligence offers substantial potential for acoustic monitoring of animals, from welfare assessment in precision livestock farming to wildlife conservation and ecological research, where vocalizations can indicate health, stress, and social states earlier and at lower cost than manual observation. However, recordings in these settings are obtained under uncontrolled conditions, including environmental noise, reverberation, overlapping calls, and sensors that degrade without notice. As a consequence, automated classification of animal vocalizations remains challenging, and the two dominant acoustic representations show complementary limitations: raw waveforms preserve temporal microstructure but degrade under clipping and reverberation, while log-Mel spectrograms capture harmonic organization but lose phase information and are sensitive to broadband noise. To address these challenges, we propose Uncertainty-Aware Fusion (UAF), a dual-stream framework that estimates Gaussian uncertainty for each representation and fuses them via uncertainty weighting. This mechanism assigns greater weight to the more confident representation with no reliability labels required. In a cross-species, identity-based evaluation excluding all individuals seen during training, UAF (mean pooling) achieves 59.4\% accuracy / 39.7\% macro F1 on the 17-class SoundWel pig vocalization benchmark and 73.1\% accuracy / 71.5\% macro F1 on the 3-class DogBark dataset, outperforming static-concatenation fusion by 15.7\% and 20.4\% relative macro F1, respectively. Ablations over four temporal aggregation strategies show that uncertainty fusion, rather than the temporal characteristics of animal calls, is the primary driver of the performance gain.
\end{abstract}




\end{frontmatter}


\section{Introduction}
\label{intro}
Precision livestock farming employs sensor-based monitoring to assess health, stress, and welfare, often at the level of the individual animal \citep{Berckmans2017}. Acoustic sensing is a widely used modality, since it requires no animal contact, covers a pen from a fixed position, and adds comparatively little hardware per monitored unit \citep{coutant2024scoping, shorten2021assessment, shorten2023acoustic}. Vocalizations have furthermore been associated with pain, thermal discomfort, hunger, and social conflict \citep{MANTEUFFEL2004163}. In recent years, deep learning-based approaches trained on raw audio or spectrogram input have reached high accuracy on curated datasets, in welfare assessment \citep{coutant2024scoping} as well as in wildlife conservation \citep{kershenbaum2025automatic} and ecological research \citep{penar2020applications}. This performance is largely obtained under favorable acquisition conditions and is often not reproduced in deployment, where signal quality is a limiting factor: recordings contain machinery and environmental noise, reverberation from hard surfaces, and overlapping calls from multiple animals, and sensors degrade gradually without this being detected at the time of recording. Since such predictions inform subsequent management decisions, quantifying the reliability of a prediction is of interest in addition to  classification performance.

Generally, there are two dominant acoustic representations, namely raw time-domain waveforms and log-Mel frequency spectrograms. While processing waveform signals preserves temporal microstructure and phase, it is sensitive to clipping and reverberation \citep{baevski2020wav2vec}. On the other hand, spectrogram-based models capture harmonic and frequency organization but discard phase and degrade under broadband noise \citep{stowell2022computational}. Therefore, neither representation is uniformly superior because the relative reliability shifts with recording conditions, which is a practically unavoidable reality in uncontrolled environments and field deployments. Existing dual-representation approaches \citep{kong2020panns, liang2022waveform} demonstrate that combining waveform and spectrogram features consistently outperforms either stream alone. However, these methods usually employ \emph{static fusion}: both representations receive fixed weights regardless of their reliability for a given input. Under the variable acoustic conditions of real-world animal monitoring, static weights propagate noise from the degraded representation rather than suppressing it. Although uncertainty-aware fusion has been explored in audio-visual human emotion recognition \citep{tellamekala2023cold, xie2024trustworthy}, no prior work has adapted reliability-driven weighting to audio dual-representation fusion for bioacoustic classification. This is challenging because, unlike audio-visual fusion, both modalities originate from the same acoustic signal, requiring the model to determine which representation should be trusted in each step in time.

We address this gap with \textbf{Uncertainty-Aware Fusion (UAF)}, a framework that dynamically weights waveform and spectrogram contributions using per-utterance latent uncertainty estimates (Fig.~\ref{fig:model}). First, modality-specific ResNet-18 encoders map each input to a temporal feature sequence, and a temporal aggregation module produces a single utterance-level embedding per modality. An uncertainty head then parameterizes a per-utterance diagonal Gaussian $\mathcal{N}(\boldsymbol{\mu}_m, \boldsymbol{\sigma}_m^2)$ for each modality $m \in \{A, F\}$, and dimension-wise uncertainty weights fuse the two means: each latent dimension is dominated by whichever modality is more confident for that input. Training optimizes a multi-term objective that learns from three streams, i.e., waveform, spectrogram, and a fusion branch. We evaluate on two publicly available bioacoustic benchmark datasets, i.e., SoundWel, a 17-class pig welfare vocalization \citep{briefer2022classification} and DogBark, a dataset for context-dependent dog bark classification \citep{yin2004barking}. 
In summary, our contributions are threefold:
\begin{itemize}
    \item We propose a dual-representation architecture for bioacoustic classification that processes raw waveform and log-Mel spectrogram inputs in parallel, capturing temporal and frequency cues that neither representation provides on its own.
    
    \item We employ an uncertainty-aware fusion mechanism that estimates the reliability of each representation at every time step and weights their contributions accordingly. This directly targets a practical problem in real-world and barn recordings, where noise, clipping, or occlusion can degrade one representation while leaving the other intact.
    
    \item We evaluate our framework on two bioacoustic benchmarks, SoundWel (pig, 17 classes) and DogBark (dog, 3 classes), and report ablation studies that show the gain from uncertainty-aware fusion relative to static concatenation baselines.
\end{itemize}

\section{Related Work}

\subsection{Acoustic Representations in Animal Bioacoustics}

The time-domain waveform and the frequency spectrogram are two different representations of acoustic signals that can be used for behavioral classification of animals. Early approaches relied on hand-crafted features for specific species and monitoring objectives \citep{coutant2024scoping, mcloughlin2019automated}. These features are grouped into three types: time-domain descriptors such as call duration and inter-call intervals; frequency-domain descriptors including fundamental frequency, formants, and spectral centroid; and energy-related measures such as amplitude and signal-to-noise ratio. Dynamic features such as jitter and shimmer capture temporal micro-variations, while derived representations such as Mel-Frequency Cepstral Coefficients (MFCCs) and their temporal derivatives have become common in automatic classification systems \citep{li2024recent}. For instance, pig distress calls exhibit higher pitch, longer duration, and noisier spectra, whereas chicken alarm calls are typically short, high-pitched bursts \citep{coutant2024scoping}. While hand-crafted features have enabled considerable progress in livestock monitoring \citep{mcloughlin2019automated} and wildlife conservation, they require substantial domain expertise, may not generalize well across recording conditions or species, and can miss subtle, nonlinear interactions between temporal and spectral information. These limitations have motivated a shift toward learned representations via deep neural networks; however, whether time-domain waveforms, frequency-domain spectrograms, or both yield the best learned features remains an open question with distinct trade-offs in information preservation, computational cost, and noise robustness.

\subsection{Spectrogram-based Deep Learning Models}

Spectrogram-based approaches have emerged as the dominant paradigm in bioacoustic deep learning, converting raw audio signals into time-frequency representations such as mel-spectrograms, MFCCs, or delta-mel features before applying, e.g., convolutional neural networks (CNNs) \citep{stowell2022computational, xie2024sound}. By treating spectrograms as images, computer vision approaches can learn hierarchical spatial patterns that capture both temporal and spectral dynamics, consistently outperforming traditional classifiers based on hand-crafted features in comparative studies \citep{ghani2023global, bergler2022animal}. Hybrid architectures that incorporate recurrent units, e.g., CNN--LSTM or CNN--GRU, further enhance performance by modeling temporal dependencies across spectrogram frames \citep{gupta2021comparing}. More recently, transformer-based models that treat the log-mel spectrogram as a sequence of 2-D patches, such as the Audio Spectrogram Transformer (AST) \citep{gong2021ast}, achieve state-of-the-art results on large-scale audio benchmarks. The AVES (Animal Vocalization Encoder based on Self-Supervision) model \citep{hagiwara2023aves} extends this paradigm to animal vocalizations via self-supervised transformer pre-training, yielding strong transfer performance on bioacoustic classification.

These spectrogram-based models have been successfully applied across diverse bioacoustic domains. In livestock applications, automated classification systems using multiple CNN backbones (e.g., AlexNet, ResNet, EfficientNet) have demonstrated high accuracy for dog vocalizations \citep{gomezarmenta2024dog}, while emotion recognition systems show promise for farm animal welfare monitoring \citep{briefer2022classification}. For wildlife monitoring, large-scale species classification of birds and insects has leveraged CNNs to capture discriminative frequency patterns across taxa \citep{nanni2020ensemble, stowell2022computational}, while in marine bioacoustics, deep learning models have been developed for detecting baleen whale vocalizations \citep{schall2024deep}. Recent frameworks such as ANIMAL-SPOT \citep{bergler2022animal} provide end-to-end pipelines for detection and classification directly from spectrograms, showing strong generalization across multiple species and call types.

Despite these empirical successes, spectrogram-based methods have known limitations. The time-frequency transformation inherently discards phase information, which may contain discriminative temporal microstructures relevant for certain call types. Model performance is sensitive to spectrogram resolution and preprocessing choices such as window size, hop length, and frequency binning, requiring careful tuning for each application. Furthermore, performance often degrades when deployed in real-world scenarios that differ from training conditions, particularly under varying noise profiles or recording equipment \citep{stowell2019birdaudio, shorten2023acoustic}. These robustness challenges suggest that relying solely on frequency-domain representations could be insufficient for generalizable bioacoustic monitoring systems.

\subsection{Raw Waveform and Hybrid Audio Models}

Motivated by the potential information loss in spectrogram conversion, raw waveform modeling has gained attention as a complementary representation strategy. Raw waveform approaches often apply convolutional filters directly to time-domain audio signals, allowing the network to learn discriminative temporal features without explicit frequency transformation \citep{tokozume2017envnet}. Self-supervised pre-training on raw waveforms---exemplified by wav2vec~2.0 \citep{baevski2020wav2vec} and HuBERT \citep{hsu2021hubert}---has demonstrated that contextual representations learned from large unlabeled audio corpora transfer effectively to downstream classification tasks. Large-scale pretrained models such as PANNs (Pretrained Audio Neural Networks) further show that fusing waveform- and spectrogram-level features improves audio pattern recognition. The Wavegram-Logmel-CNN variant in PANNs concatenates features from both domains and achieves better performance than either branch alone \citep{kong2020panns}. The theoretical advantage of waveform models lies in preserving complete signal information, including phase relationships and fine-grained temporal structures that may be smoothed or lost during spectrogram computation.

In bioacoustics, however, raw waveform modeling remains relatively underexplored compared to spectrogram-based approaches. Bravo~Sanchez et al.\ demonstrate that SincNet-based waveform architectures can classify avian calls without spectrograms, establishing the viability of end-to-end waveform learning for bioacoustic tasks \citep{bravoSanchez2021bioacoustic}. Nevertheless, bioacoustic applications continue to face unique challenges, including data scarcity, high intra-species variability, and diverse recording conditions. More critically, although both representations offer complementary information, existing multi-representation approaches such as PANNs \citep{kong2020panns} as well as the dual-stream waveform and log-mel fusion of \citet{liang2022waveform} rely on static, fixed-weight concatenation that does not adapt when one representation becomes less reliable under specific conditions. The key unresolved challenge is not whether to combine both representations but how to dynamically assess and respond to their relative reliability.

\subsection{Multimodal Fusion Strategies}

The recognition that single representations may be insufficient has led to growing interest in multimodal integration, both across different sensor types and representations of the same audio signal. Baltru\v{s}aitis et al. provide the foundational taxonomy of multimodal learning strategies, distinguishing early fusion, late fusion, and intermediate fusion, and highlighting representation learning and fusion as the two central challenges \citep{baltrusaitis2019multimodal}. In livestock monitoring, recent systems have combined acoustic signals with accelerometry and inertial sensors to improve the classification of behavioral states and welfare indicators \citep{vu2024mmcows, ferrero2025multihead}. These multimodal systems leverage complementary information from different sensing modalities to provide more robust predictions in noisy farm environments, where visual occlusion, acoustic interference, or sensor failures can degrade individual channels.

However, most existing fusion strategies employ simple integration methods that assume equal reliability across modalities. Early fusion approaches concatenate raw features or low-level representations before classification, while late fusion combines independent predictions from modality-specific classifiers through averaging or voting. Intermediate strategies extract separate embeddings from each modality and concatenate them before the final classification layers \citep{baltrusaitis2019multimodal}. These methods treat all modalities equally informative at all times, assigning fixed or learned but static weights that do not adapt to temporal variations in signal quality. Peng et al.\ show empirically that joint multimodal training suffers from a modality-imbalance problem and propose on-the-fly gradient modulation as a solution \citep{peng2022balanced}.

This assumption of uniform reliability is problematic in real-world deployments, where waveform and spectrogram representations show input-dependent failure modes. Existing fusion methods amplify the noise from a degraded representation rather than suppress it, motivating reliability-adaptive weighting.

\subsection{Uncertainty and Reliability Modeling in Multimodal Learning}

The concept of uncertainty-aware fusion, where modality contributions are dynamically weighted based on estimated reliability, has roots in Bayesian deep learning. Gal and Ghahramani showed that Monte Carlo dropout provides a computationally tractable approximation to Bayesian inference, yielding epistemic uncertainty estimates at test time without modifying the training procedure \citep{gal2016dropout}, while Kendall and Gal formalized the canonical decomposition into aleatoric uncertainty (irreducible data noise) and epistemic uncertainty (model uncertainty reducible with more data) \citep{kendall2017uncertainties}. In the multimodal domain, \citet{subedar2019uncertainty} applied deep Bayesian variational inference to audio-visual activity recognition and demonstrated that per-modality uncertainty estimates improve robustness when one sensor stream is corrupted.  \citet{han2022trusted} extended uncertainty-driven fusion to arbitrary numbers of views via Dempster-Shafer evidence fusion, enabling reliable classification under missing or degraded inputs. In multimodal emotion recognition, the COLD Fusion framework \citep{tellamekala2023cold} introduced a principled approach by learning latent distributions over unimodal temporal contexts and constraining their variance to represent how informative each modality is toward the prediction task. By imposing calibration and ordinal ranking constraints on modality-wise uncertainty estimates, COLD Fusion enables adaptive weighting that favors more reliable modalities at each time step.

\section{Methods}
The core idea of the proposed Uncertainty-Aware Fusion (UAF) framework is to replace static fusion weights for time-domain waveform and spectrogram representations with adaptive weights estimated directly from the learned features. This enables the model to quantify modality-specific uncertainty and suppress less reliable representations without requiring explicit reliability annotations. We first formulate the classification task and establish the notation used throughout the paper. We then describe the proposed framework in detail, including the overall architecture, unimodal waveform and spectrogram encoders, temporal aggregation, uncertainty estimation, the adaptive fusion strategy, and the corresponding training objective.

\subsection{Problem Definition}
The task we address in this paper is to classify animal vocalizations into $K$ discrete behavioral categories. Given an audio recording, two complementary representations are extracted: a raw waveform $x_A \in \mathbb{R}^{L}$ and a three-channel stacked spectrogram $x_F \in \mathbb{R}^{C \times H \times W}$, where $L$ is the waveform length, $C=3$ channels encode log-Mel, $\Delta$-Mel, and $\Delta\Delta$-Mel coefficients, and $H \times W$ is the spectrogram spatial resolution \citep{young2002htk, takahashi2016deep}. We consider three-channel spectrogram input rather than the one-channel log-Mel because the derivative channels capture short-term spectral dynamics (onset, offset, and formant-transition cues) that the log-Mel spectrogram discards \citep{young2002htk}. The goal is to learn a mapping $f : (x_A, x_F) \mapsto \hat{y} \in \{1,\ldots,K\}$ that minimizes classification error while producing calibrated confidence estimates. 

\subsection{Architecture Overview}
\label{sec:overview}

\begin{sidewaysfigure}
    \centering
    \includegraphics[width=0.98\linewidth]{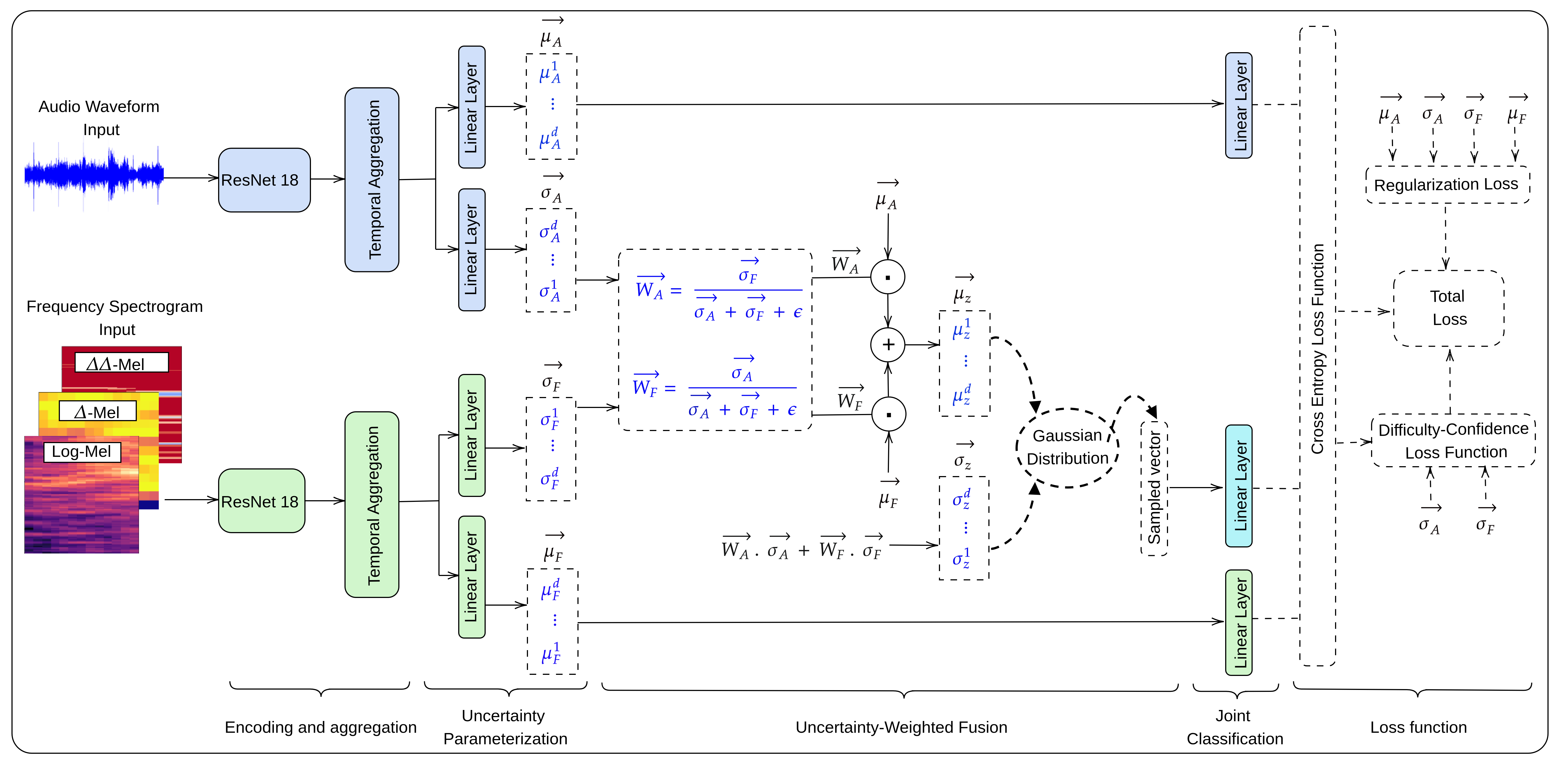}
    \caption{Overview of the proposed Uncertainty-Aware Fusion (UAF) framework for animal vocalization classification. Audio waveform and frequency specgrogram are encoded separately and temporally aggregated. The resulting embeddings are mapped to Gaussian means and variances, and the uncertainty-aware fusion module combines the two modalities into a single fused distribution by assigning higher weights to the more certain modality. A vector sampled from this fused distribution, together with two auxiliary unimodal predictions, is used for the final classification. A difficulty-confidence loss couples each modality's uncertainty to how hard its predictions are, while a regularization term keeps the uncertainties feasible for an end-to-end train.}
    \label{fig:model}
\end{sidewaysfigure}

Figure~\ref{fig:model} illustrates the proposed Uncertainty-Aware Fusion (UAF) framework, which we realize in four steps. First, we perform \textbf{modality encoding} by passing each modality through a dedicated ResNet-18 encoder, producing a temporal sequence of $T$ feature vectors of dimension $C_e=512$.  These features are compressed by a temporal aggregation module into a single utterance-level embedding $\mathbf{h}_m \in \mathbb{R}^{C_e}$ (see Sections \ref{sec:encoder} and \ref{sec:temporal}). Then, we perform \textbf{uncertainty parameterization} by mapping each embedding to the mean and variance of a diagonal Gaussian, $q_m(z \mid \mathbf{h}_m) = \mathcal{N}(\boldsymbol{\mu}_m, \mathrm{diag}(\boldsymbol{\sigma}_m^2))$ for $m \in {A,F}$, where $A$ and $F$ denote the waveform and spectrogram modalities. We use the predicted variance as an implicit measure of modality reliability, enabling the model to adaptively weight each modality without requiring explicit reliability annotations (see Section \ref{sec:unc_head}). Next, in \textbf{uncertainty-weighted fusion} step, we combine these Gaussians dimension-wise, giving more weight to whichever modality has lower variance in each latent dimension, and obtain a fused latent sample $\mathbf{z} \in \mathbb{R}^{D_z}$ (see Section \ref{sec:fusion}). Finally, we do a \textbf{joint classification} by applying a linear classifier to fused representation $\mathbf{z}$ to obtain the final prediction, while two auxiliary classifiers are applied to $\boldsymbol{\mu}_A$ and $\boldsymbol{\mu}_F$ to provide modality-specific supervision. We train the whole model with a loss function inspired by \citep{tellamekala2023cold} and \citep{xie2024trustworthy} that couples the predicted variances with the observed classification difficulty (see Section \ref{sec:loss}).

\subsection{Dual Audio Encoder}
\label{sec:encoder}
We use separate encoders for the raw waveform and the spectrogram:
\paragraph{Waveform encoder}
We use a custom 1-D ResNet-18 to map a single-channel waveform segment of length $L$ (the number of raw audio samples) to a temporal feature map:
\begin{equation}
\mathbf{E}_A : \mathbb{R}^{L} \to \mathbb{R}^{C_e \times T}, \qquad T = \lfloor L / 32 \rfloor,
\label{eq:wav_encoder}
\end{equation}
where $C_e$ is the number of feature channels the encoder produces at each time step and $T$ is the resulting sequence length (the number of temporal feature vectors handed to the aggregation module, see Section \ref{sec:temporal}). The $32$-fold downsampling from $L$ to $T$ accumulates over a stride-2 stem (conv\,$7{\times}1$ $+$ max-pool) followed by three stride-2 residual stages. The values of $C_e$, $L$, or $T$ are not intrinsic to the fusion mechanism: $C_e = 512$ reflects the standard ResNet-18 backbone, $L = 6{,}400$ results from sampling a raw audio waveform at $16$~kHz, and the sampling rate (400~ms) was chosen to match the typical duration of a single vocalization event in our target datasets. Employing a different backbone, segment length, or sampling rate would rescale these values without altering the underlying uncertainty-aware fusion method.

\paragraph{Spectrogram encoder}
We use a standard ResNet-18 \citep{he2016deep} to produce frequency features:  $\mathbf{E}_F : \mathbb{R}^{C \times H \times W} \to \mathbb{R}^{C_e \times T}$. This means that the backbone maps $(B, C, H, W) \to (B, C_e, H/32, W/32)$, then an average pooling collapses the frequency axis, and linear interpolation aligns the temporal axis to $T$. With $H = W = 128$ the backbone yields a $4 \times 4$ spatial map after frequency pooling. There are four distinct temporal observations, which are linearly interpolated to $T = 200$ to match the waveform encoder's output shape $(B, C_e, T)$. 

\subsection{Temporal Context Aggregation}
\label{sec:temporal}

Each encoder sequence $\{e_{m,t}\}_{t=1}^{T} \subset \mathbb{R}^{C_e}$ is aggregated into a single utterance-level embedding $\mathbf{h}_m \in \mathbb{R}^{D_h}$. To yield this embedding, our proposed UAF model applies global average pooling:
\begin{equation}
\mathbf{h}_m = \frac{1}{T}\sum_{t=1}^{T} e_{m,t}, \qquad D_h = C_e = 512.
\label{eq:mean_pool}
\end{equation}
We additionally evaluate three alternative aggregation modules as ablations of this design choice:
\begin{itemize}
\item \textit{GRU}: A multi-layer GRU processes the sequence in temporal order, and $\mathbf{h}_m$ is taken as the final hidden state of the last recurrent layer.
\item \textit{Transformer}: A Pre-Layer Normalization Transformer encoder with learnable positional embeddings operates on the time step sequence, and the output is mean-pooled to $\mathbf{h}_m \in \mathbb{R}^{C_e}$.
\item \textit{Attention pooling}: A two-layer MLP processes the embeddings of each time step, and a softmax over the sequence turns the output into attention weights such that the time steps that are more informative contribute more.

\end{itemize}
For all variants, $\mathbf{h}_m$ is further processed by the downstream uncertainty head and fusion module, isolating temporal aggregation as the critical architectural design in the ablation.

\subsection{Gaussian Uncertainty Parameterization}
\label{sec:unc_head}

The gaussian uncertainty parameterization is the major part where our approach differs from a conventional dual-stream classifier. Instead of treating each modality's utterance embedding $\mathbf{h}_m$ as a fixed point, an uncertainty head maps it to a distribution, producing a mean $\boldsymbol{\mu}_m$  and a variance $\boldsymbol{\sigma}_m^2$. The mean plays the role of a conventional embedding, while the variance indicates the model's confidence in that estimate. Therefore, the variance should be small along dimensions where a modality is clear and unambiguous, and it should be large along dimensions where the input is noisy or degraded. As there are no reliability labels, the association between uncertainty and input quality is learned using a loss function inspired by \citep{tellamekala2023cold} and \citep{xie2024trustworthy}. We explain the loss function and its empirical evaluation in Sections \ref{sec:loss} and \ref{sec:calibration}, respectively.

We use two independent linear layers to yield $\boldsymbol{\mu}_m$ and $\boldsymbol{\sigma}_m$ from the embeddings of the above-described waveform and spectrogram encoders:
\begin{equation}
\boldsymbol{\mu}_m = W_{\mu}^{(m)}\mathbf{h}_m, \qquad
\boldsymbol{\sigma}_m = \mathrm{softplus}\!\left(W_{\sigma}^{(m)}\mathbf{h}_m\right) + \varepsilon,
\label{eq:unc_head}
\end{equation}
where $W_{\mu}^{(m)}, W_{\sigma}^{(m)} \in \mathbb{R}^{D_z \times D_h}$ are learnable projections. Softplus keeps $\boldsymbol{\sigma}_m$  positive while remaining differentiable everywhere, such that it can be trained end-to-end with standard backpropagation. Adding the small constant $\varepsilon = 10^{-6}$ guarantees $\boldsymbol{\sigma}_m$ never reaches exactly zero, which protects the fusion rule given in Section \ref{sec:fusion} from dividing by zero. We consider the dimensionality of $\boldsymbol{\mu}_m$ and $\boldsymbol{\sigma}_m$ as a hyperparameter ($D_z$) and optimize it using Bayesian optimization leveraging the Python framework Optuna  \citep{akiba2019optuna} (see Section \ref{subsec:training}).

\subsection{Dimension-wise Uncertainty Fusion}
\label{sec:fusion}

Given the Gaussian parameters $(\boldsymbol{\mu}_A,\boldsymbol{\sigma}_A)$ and $(\boldsymbol{\mu}_F,\boldsymbol{\sigma}_F)$, fusion weights are computed independently for each latent dimension $d \in \{1,\ldots,D_z\}$:
\begin{equation}
w_A(d) = \frac{\sigma_F(d)}{\sigma_A(d) + \sigma_F(d)}, \qquad
w_F(d) = \frac{\sigma_A(d)}{\sigma_A(d) + \sigma_F(d)}.
\label{eq:fusion_weights}
\end{equation}
Consequently, a dimension in which the spectrogram is highly uncertain ($\sigma_F(d) \gg \sigma_A(d)$) receives proportionally more weight from the waveform, and vice versa. Stacking these per-dimension weights into vectors $\mathbf{w}_A, \mathbf{w}_F \in \mathbb{R}^{D_z}$ (with $\mathbf{w}_A + \mathbf{w}_F = \mathbf{1}$), the fused Gaussian parameters are defined as:
\begin{align}
\boldsymbol{\mu}_z &= \mathbf{w}_A \odot \boldsymbol{\mu}_A + \mathbf{w}_F \odot \boldsymbol{\mu}_F,
\label{eq:fused_mu} \\
\boldsymbol{\sigma}_z &= \mathbf{w}_A \odot \boldsymbol{\sigma}_A + \mathbf{w}_F \odot \boldsymbol{\sigma}_F.
\label{eq:fused_sigma}
\end{align}
During training, a latent sample is drawn via the reparameterization trick:
\begin{equation}
\mathbf{z} = \boldsymbol{\mu}_z + \boldsymbol{\sigma}_z \odot \boldsymbol{\varepsilon}, \quad
\boldsymbol{\varepsilon} \sim \mathcal{N}(\mathbf{0}, \mathbf{I}).
\label{eq:reparam}
\end{equation}
At inference, we set $\mathbf{z} = \boldsymbol{\mu}_z$. The vector $\mathbf{z}$ is then further processed by a dropout-regularized linear layer to produce class logits $\hat{\ell}_z = W_c\mathbf{z} \in \mathbb{R}^K$.

\subsection{Training Objective}
\label{sec:loss}

Inspired by \citet{tellamekala2023cold} and  \citet{xie2024trustworthy}, we design a novel loss function that supervises classification from the fused and unimodal predictions, and simultaneously learns the uncertainty to help a dynamic fusion based on the confidence of the modalities. In the following, we explain each component of this loss function.

\paragraph{Classification loss}
Each unimodal stream shown in Fig.~\ref{fig:model} has an independent linear head on top of $\boldsymbol{\mu}_A$ and $\boldsymbol{\mu}_F$, allowing it to learn the class label directly from that modality. The fused branch (logits from $\mathbf{z}$) is additionally supervised using the same class label. Together, the three branches form the classification loss:
\begin{equation}
\mathcal{L}_{\mathrm{cls}} = \frac{1}{3}\Bigl[\mathcal{H}(\hat{\ell}_z,\, y) + \mathcal{H}(\hat{\ell}_A,\, y) + \mathcal{H}(\hat{\ell}_F,\, y)\Bigr],
\label{eq:l_cls}
\end{equation}
where $\mathcal{H}(\cdot,y)$ denotes cross-entropy with the ground-truth label $y$, and $\hat{\ell}_m = W_{m}\boldsymbol{\mu}_m$. During training, $\hat{\ell}_z$ is computed from the reparameterized sample, see Eq.~\ref{eq:reparam}, while in the inference phase, the unimodal classifiers receive means $\boldsymbol{\mu}_A$ and $\boldsymbol{\mu}_F$ without the reparameterization noise.

\paragraph{Difficulty-Confidence loss}
\label{par:DC_loss}
The fusion process requires uncertainty scores. However, no ground-truth labels for uncertainty are available. Therefore, we compare within each mini-batch of $B$ samples how hard a sample actually was to classify against how confident the model was about its prediction. This allows the model to learn meaningful uncertainty estimates from relative comparisons within a batch.

For each modality $m \in \{A,F\}$ and sample $i$, we compute a difficulty score (the sample's own cross-entropy loss) and a confidence score (the inverse of its predicted uncertainty):
\begin{equation}
d_m^{(i)} = \mathcal{H}(\hat{\ell}_m^{(i)},\, y^{(i)}), \qquad
s_m^{(i)} = \frac{1}{\|\boldsymbol{\sigma}_m^{(i)}\|_2 + \varepsilon}.
\label{eq:d_s}
\end{equation}
Both scores are then normalized via a softmax with temperature $\tau$ over all $B$ samples into two probability distributions describing how difficult and how confident each sample is \emph{relative to the rest of the batch}:
\begin{equation}
\mathbf{p}_D^m = \mathrm{softmax}\!\left(\mathbf{d}_m\,/\,\tau\right), \qquad
\mathbf{p}_S^m = \mathrm{softmax}\!\left(\mathbf{s}_m\,/\,\tau\right).
\label{eq:batch_dists}
\end{equation}
Matching these two distributions via a symmetric KL divergence encourages samples that rank as relatively difficult under $\mathbf{p}_D^m$ to also rank as relatively confident under $\mathbf{p}_S^m$, which corresponds to lower $\boldsymbol{\sigma}_m$. In contrast, easy samples are encouraged to have lower confidence (higher $\boldsymbol{\sigma}_m$). Thus, the loss couples relative difficulty to relative confidence. The results in Section \ref{sec:calibration} show the correlation between $\|\sigma_z\|$ and the misclassification error. For each modality $m$, this correlation is imposed with the symmetric KL divergence:
\begin{equation}
\mathcal{L}_{\mathrm{DC},m} = \mathrm{KL}\!\left(\mathbf{p}_D^m \,\|\, \mathbf{p}_S^m\right) + \mathrm{KL}\!\left(\mathbf{p}_S^m \,\|\, \mathbf{p}_D^m\right), \quad m \in \{A, F\}.
\label{eq:co_intra}
\end{equation}
To apply the same idea jointly (as a single loss term) across both modalities, we interleave the modalities' difficulty and confidence scores into $2B$-length vectors:
\begin{equation}
\mathbf{d}_{AF} = \bigl[d_A^{(1)},\, d_F^{(1)},\,\ldots,\, d_A^{(B)},\, d_F^{(B)}\bigr], \quad
\mathbf{s}_{AF} = \bigl[s_A^{(1)},\, s_F^{(1)},\,\ldots,\, s_A^{(B)},\, s_F^{(B)}\bigr].
\label{eq:d_af}
\end{equation}
The same symmetric KL penalty is then applied to this combined set of $2B$ samples:
\begin{equation}
\mathcal{L}_{\mathrm{DC},AF} = \mathrm{KL}\!\left(\mathbf{p}_D^{AF} \,\|\, \mathbf{p}_S^{AF}\right) + \mathrm{KL}\!\left(\mathbf{p}_S^{AF} \,\|\, \mathbf{p}_D^{AF}\right).
\label{eq:co_cross}
\end{equation}

\paragraph{Variance regularizer}
To prevent variance collapse (the model pushing $\boldsymbol{\sigma}_m \to 0$ to suppress noisy gradients), we penalize departure from an isotropic standard normal prior \citep{tellamekala2023cold, xie2024trustworthy}:
\begin{equation}
\begin{split}
\mathcal{L}_R &= \sum_{m \in \{A,F\}} \mathrm{KL}\!\left(\mathcal{N}(\boldsymbol{\mu}_m, \boldsymbol{\sigma}_m^2)\,\big\|\,\mathcal{N}(\mathbf{0}, \mathbf{I})\right) \\
&= -\frac{1}{2}\sum_m \sum_{d=1}^{D_z}\Bigl[1 + \log \sigma_{m,d}^2 - \mu_{m,d}^2 - \sigma_{m,d}^2\Bigr],
\end{split}
\label{eq:var_regu}
\end{equation}
averaged over the batch.

The total training objective composed of the above-described components is finally formulated as:
\begin{equation}
\mathcal{L} = \lambda_{\mathrm{cls}}\,\mathcal{L}_{\mathrm{cls}}
            + \lambda_{\mathrm{DC},A}\,\mathcal{L}_{\mathrm{DC},A}
            + \lambda_{\mathrm{DC},F}\,\mathcal{L}_{\mathrm{DC},F}
            + \lambda_{\mathrm{DC},AF}\,\mathcal{L}_{\mathrm{DC},AF}
            + \lambda_R\,\mathcal{L}_R,
\label{eq:total_loss}
\end{equation}
where default weights are set empirically to  $\lambda_{\mathrm{cls}}=1$, $\lambda_{\mathrm{DC},A}=\lambda_{\mathrm{DC},F}=\lambda_{\mathrm{DC},AF}=10^{-3}$, $\lambda_R=10^{-4}$.

\section{Experimental Setup}

\subsection{Datasets and Data Splits}

We evaluate the proposed uncertainty-aware fusion framework on two publicly available datasets representing (i) different animal vocalization classification tasks to assess cross-species generalization and (ii) different task complexities: pig calls and dog barks context classification.

\paragraph{SoundWel Dataset} The SoundWel dataset \citep{briefer2022classification} is a multi-institutional collection of pig vocalizations under the SoundWel research program. It contains 6,887 labeled audio clips recorded by six independent research teams across Europe (ETHZ, FBN, IASPA, IASPB, IASPC, NMBU), covering three age groups (piglet, weaner, growing/finishing) and 17 behavioral context categories spanning welfare-negative states (e.g.\ isolation, castration, crushing), welfare-positive states (e.g.\ enriched, reunion), and nursing-related behaviors. The dataset is severely class-imbalanced: For instance, isolation accounts for 29.7\% of samples, while surprise accounts for only 0.2\%. Individual clip durations have a mean of 0.36~s and a maximum of 3.6~s.

\paragraph{DogBark Dataset} The DogBark dataset \citep{yin2004barking} contains dog vocalizations from ten individual dogs spanning six breeds, annotated across three behavioral contexts: aggression (14.3\%), contact (55.6\%), and play (30.2\%). The original dataset comprises 4{,}672 samples; however, only 693 labeled recordings are publicly available.

\paragraph{Data Splits} A random stratified split would cause the recordings of the same individual to appear in both training and test sets, encouraging the model to exploit individual-specific vocal fingerprints rather than context-discriminative features. This is known as data leakage between train and test sets, which unrealistically inflates the model's performance during the test phase \citep{bernett2024guiding}. Instead, in this work, we considered identity-aware splits for both datasets. However, results under a random stratified split are also provided in Table~\ref{tab:sup_random_split} (\ref{sec:supp}) for comparison with the previously published result of \citet{briefer2022classification}.

The \textbf{SoundWel} dataset does not provide an identity train-val-test split, and only a limited number of samples have a pig ID. To avoid data leakage, we used the available metadata, i.e., filename patterns encoding per-animal or per-group identifiers (pen IDs, animal IDs, sow IDs, or litter numbers depending on team), recording year to disambiguate cohorts across experimental sessions, and recording team as a coarse grouping variable to prevent cross-team leakage. Recordings for which no parseable identity could be inferred (2,398 recordings, 34.8\% of the dataset) are assigned to the training set such that the model could benefit from learning on more samples without the risk of data leakage. The resulting split is: train~5,452 (79.2\%) $|$ val~677 (9.8\%) $|$ test~758 (11.0\%). All six recording teams contribute to every subset, and 13 of the 17 classes appear in the test set. The four absent classes in the test set (BeforeNursing, Fighting, Huddling, Run) have no recordings with parseable pig identities. Therefore, the macro-averaged F1-score is computed over the 13 classes present in the test partition.

For the \textbf{DogBark} dataset, we ensure that individual dogs are assigned exclusively to one partition. The test set consists of Mac (German shorthaired pointer, male) and Zoe (Australian cattle dog, female), providing cross-breed and cross-sex generalization with all three behavioral classes represented. The validation set consists of Keri and Louie; the remaining six dogs form the training set. This data split leads to 458 train (66.1\%), 75 val  (10.8\%), and 160 test (23.1\%) samples. The class distribution is nearly identical across partitions, making the macro-averaged F1-score a fair evaluation metric. 

\subsection{Preprocessing and Data Augmentation}

\paragraph{Audio Preprocessing} All audio recordings are resampled to 16 kHz sampling rate, which captures the relevant acoustic bandwidth for both pig and dog vocalizations (predominantly below 8 kHz) while reducing computational cost compared to higher sampling rates. This sampling rate aligns with standard practices in animal bioacoustics and enables stable training \citep{stowell2022computational}.

For the waveform input ($x_A$), raw audio signals are peak-normalized by dividing by the absolute maximum amplitude, mapping each clip to a range of $[-1, 1]$. For the spectrogram input ($x_F$), we compute a three-channel time-frequency representation using a Mel-filterbank applied to the power spectrogram (STFT with FFT size $n_{\text{fft}} = 1024$, hop length $= 256$ samples (16~ms), Hann window, and power $= 2$). We extract 128 Mel-frequency bands spanning 0--8~kHz and convert to decibels (dB) to obtain the log-power Mel spectrogram. We then compute its first-order ($\Delta$-Mel) and second-order temporal derivative ($\Delta\Delta$-Mel). These three representations are stacked as input channels and resized to $128 \times 128$ pixels. Each channel is independently min-max normalized to $[0, 1]$.

\paragraph{Temporal Segmentation} As illustrated in Fig.~\ref{fig:train test overview}, we use different temporal segmentation strategies during training and testing. During training, each audio recording is randomly cropped to a single 400 ms segment, which serves as an independent training sample. Recordings shorter than 400 ms are zero-padded: during training, the audio is placed at a random position within the target window; during validation and test, it is center-aligned. This random cropping acts as a form of data augmentation, exposing the model to different temporal regions of the same vocalization across training epochs and improving robustness to temporal variability.

The choice of 400 ms segments provides enough temporal context for both datasets and is motivated by bioacoustic considerations: Individual pig vocalizations have typical durations ranging from 0.3 to 1.8 seconds \citep{briefer2022classification}. Hence, a 400 ms window likely captures the most discriminative part of a call without padding artifacts for the majority of samples. For DogBark, individual bark events are also short (typically well under one second), and the 400 ms window likely captures a complete bark unit. Shorter segments risk capturing only partial vocal units, reducing temporal context, while segments longer than a single call risk mixing adjacent events or silent gaps. 

\paragraph{Data Augmentation} To improve model robustness to acoustic variability in realistic farm environments and prevent overfitting, we apply additive Gaussian noise during training. For each training sample, noise is generated from a standard normal distribution and scaled to achieve a signal-to-noise ratio (SNR) uniformly sampled from the range [-5, 15] dB. 

\subsection{Training Procedure}
\label{subsec:training}
The training process consists of two stages: hyperparameter optimization followed by a final training.

\paragraph{Stage 1: Hyperparameter Optimization} We use Bayesian optimization implemented via the Python framework Optuna \citep{akiba2019optuna} with a Tree-structured Parzen Estimator (TPE) sampler and a median pruner to run 50 trials. Each trial trains and evaluates the model using the previously described training and validation data. To prevent further class imbalance between folds for the SoundWel dataset, we did not employ cross-validation. For DogBark, we adopted the same strategy for consistency. Training runs for up to 300 epochs with early stopping (patience 20, monitored on validation macro-averaged F1-scores). The trial that yields the highest score is selected for final training. The hyperparameter search space covers: learning rate $\in [10^{-5}, 10^{-3}]$, latent dimension $D \in \{64, 128, 256\}$, softmax temperature $\tau \in [0.5, 2.0]$, batch size $\in \{16, 32, 64\}$, weight decay $\in [10^{-6}, 10^{-4}]$, classification loss weight $\lambda_{\text{cls}} \in [0.5, 2.0]$, intra-modal DC loss weights $\lambda_{\text{DC\_a}}, \lambda_{\text{DC\_f}} \in [2{\times}10^{-3}, 2{\times}10^{-1}]$, cross-modal DC loss weight $\lambda_{\text{DC\_af}} \in [2{\times}10^{-3}, 2{\times}10^{-1}]$, and variance regularization weight $\lambda_{\text{regu}} \in [2{\times}10^{-5}, 2{\times}10^{-3}]$. 

\paragraph{Stage 2: Final Model Training} Using the best hyperparameter configuration from Stage 1, the training and validation data are merged, such that the final model trains on all non-test data. No early stopping is applied; instead, we scale the number of training epochs based on the results of the hyperparameter optimization (HPO):
\begin{equation}
T_{\text{final}} = \operatorname{round}\!\left( T_{\text{hpo}} \times \frac{N_{\text{hpo}}}{N_{\text{final}}} \right),
\label{eq:epoch_scaling}
\end{equation}
where $T_{\text{hpo}}$ is the epoch at which early stopping was conducted for the best HPO trial, $N_{\text{hpo}}$ is the training-set size during HPO, and $N_{\text{final}}$ is the size of the combined training and validation data. For the final training, we use the AdamW optimizer \citep{loshchilov2017decoupled} with a linear warm-up schedule (learning rate rises from $\mathrm{lr} \times 10^{-3}$ to $\mathrm{lr}$ over 10 epochs) followed by cosine annealing to a minimum of $10^{-6}$.

The identity-constrained test partitions are rather small (160 recordings for DogBark, 758 for SoundWel). To yield a reliable performance assessment and reduce the effect of randomness in training, we repeat Stage~2 five times and provide an averaged performance over five seeds. The performance variation  is, for instance, caused by random weight initialization, per-step reparameterization noise ($z = \mu_z + \sigma_z \odot \varepsilon,\ \varepsilon \sim \mathcal{N}(0,I)$), and stochastic data augmentation. 

\subsection{Test-Time Inference and Segment Aggregation}

\begin{figure}
    \centering
    \includegraphics[width=0.98\linewidth]{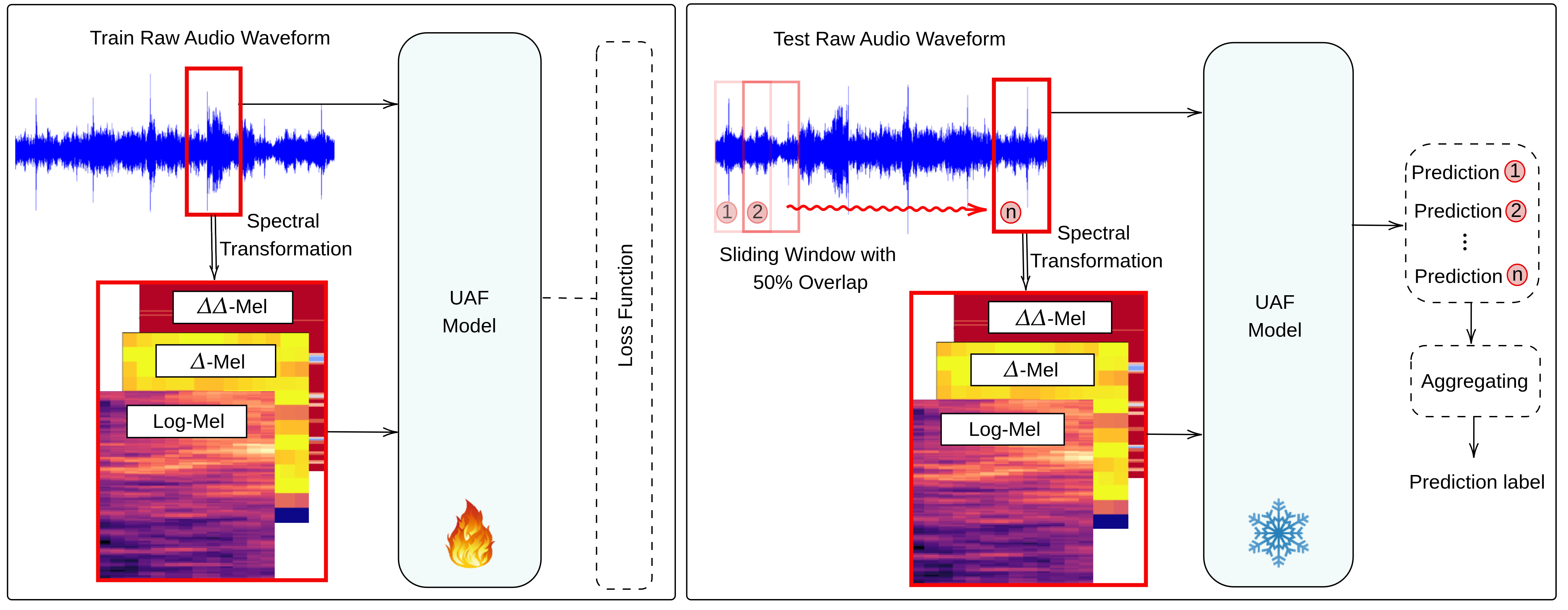}
    \caption{Training and testing procedures. Left: during training, random 400 ms crops are extracted from each recording. Right: during testing, overlapping 400 ms sliding windows with 50\% overlap are applied, and predictions are aggregated across all windows.}
    \label{fig:train test overview}
\end{figure}

At test time, audio recordings are processed at their full length rather than as single random crops, ensuring that all temporal information is utilized for classification. As shown in Fig.~\ref{fig:train test overview}, each recording is partitioned into overlapping segments of 400 ms duration with 50\% overlap, producing several chunks. Recordings shorter than 400 ms are represented by a single chunk and right-padded with zeros to the target length.

Each chunk is independently processed by the model to produce a per-chunk prediction, and the final recording-level prediction is determined by a majority vote across all chunks. We used this strategy since the overlapping chunks ensure comprehensive coverage of the recording, and the majority vote is robust to individual chunks containing noise or ambiguous features.

\subsection{Implementation Details}

All experiments were implemented with PyTorch Lightning and trained on NVIDIA RTX 6000, with 48GB VRAM using Distributed Data Parallelism and 16-bit mixed-precision arithmetic. All hyperparameters are HPO-selected independently per model variant and dataset; exact values are recorded in the experiment configuration files released with the code. Both the waveform and spectrogram encoders are trained from random initialization; no pretrained weights are used. All variants share identical backbone architectures and training conditions to ensure that performance differences reflect the fusion strategy rather than encoder capacity. The \textbf{Concatenation} baseline uses the same dual ResNet-18 encoders and mean-pooling aggregation as the proposed UAF model; the two 512-dimensional utterance embeddings are concatenated into a 1{,}024-dimensional vector and passed to a linear classification head trained with cross-entropy only. The waveform-only and spectrogram-only baselines likewise use single-branch cross-entropy. The UAF uncertainty heads ($W_\mu^{(m)}$ and $W_\sigma^{(m)}$ in Eq.~\ref{eq:unc_head}) add approximately 262K model parameters over the Concatenation baseline when $D_z=128$. All code is publicly available on GitHub at \url{https://github.com/smAIL-WS/Uncertainty-Aware-Crossmodal-Fusion-for-Classification-of-Animal-Behavior}

\section{Results and Discussion}
\label{sec:results}
We first provide a quantitative performance comparison of our proposed UAF framework on both benchmark datasets. Then, we further analyze the per-class results, the prediction calibration and confidence, as well as the fusion weight adaptation if one of the representations is corrupted or missing. Model performance is evaluated on held-out test data, with all metrics reported as mean and standard deviation over five independent training runs.

\subsection{Quantitative Performance Comparison}
\label{sec:quant}

In Table~\ref{tab:performance}, we compare seven model variants: two unimodal baselines ((i) waveform-only, (ii) spectrogram-only), (iii) simple concatenation fusion, and four uncertainty-aware fusion (UAF) configurations differing in their temporal module ((iv) mean pooling, (v) GRU, (vi) Transformer, (vii) attention pooling).

On the SoundWel dataset, the spectrogram-only baseline (57.7\% acc., 37.2\% F1) substantially outperforms the waveform-only baseline (52.2\% acc., 34.7\% F1), confirming that frequency-domain representations capture more discriminative information for the pig vocalization context. Further, simple concatenation (56.4\% acc., 34.3\% F1) underperforms the spectrogram-only on both metrics. Hence, merging a weaker waveform stream with a stronger spectrogram without accounting for their relative reliability can dilute the discriminative frequency features. All UAF variants avoid this degradation, which means the uncertainty weighting shifts contribution toward whichever modality produces more confident embeddings. UAF (mean pooling) achieves the best macro-averaged F1-score (39.7\%~$\pm$~1.3\%), while UAF (attention pooling) achieves the highest accuracy (60.3\%~$\pm$~2.5\%). Our experimental results thus indicate that the primary performance advantage on SoundWel is the avoidance of concatenation degradation ($-$2.9 F1 vs.\ spectrogram-only) rather than a large absolute gain over the best unimodal baseline. The GRU and Transformer variants are competitive but show larger run-to-run variance, which we attribute to the larger hypothesis space of these models on a dataset of limited size.

On the DogBark dataset, UAF (mean pooling) is the best-performing model across both metrics (73.1\%~$\pm$~6.9\% acc., 71.5\%~$\pm$~5.7\% F1), outperforming spectrogram-only by 4.0 percent in accuracy and 8.3 percent in F1-score, and outperforming concatenation by 10.7 and 12.1 percent, respectively. This large margin over concatenation indicates that the adaptive modality weighting of UAF provides a meaningful robustness benefit for small datasets such as DogBark. UAF (Transformer) exhibits unstable training (63.3\%~$\pm$~11.0\% acc., 51.9\%~$\pm$~22.4\% F1), with some runs converging well and others collapsing, a known limitation of Transformers when training data per class is limited. Further, UAF (attention pooling) (57.0\%~$\pm$~6.2\% acc.) underperforms the concatenation method on DogBark, suggesting that the attention pooling aggregation may overfit for DogBark with only six training identities.

\begin{table*}[t]
\centering
\footnotesize
\setlength{\tabcolsep}{4pt}
\resizebox{\textwidth}{!}{%
\begin{tabular}{lcccc}
\toprule
\textbf{Model}
  & \multicolumn{2}{c}{\textbf{SoundWel} (pig-identity split, 13/17 classes in test)}
  & \multicolumn{2}{c}{\textbf{DogBark} (3 classes, dog-identity split)} \\
\cmidrule(lr){2-3}\cmidrule(lr){4-5}
 & Acc.\ (\%) & Macro F1 (\%) & Acc.\ (\%) & Macro F1 (\%) \\
\midrule
Waveform-only          & $52.2 \pm 2.0$  & $34.7 \pm 1.2$  & $53.6 \pm 2.2$  & $44.0 \pm 11.2$ \\
Spectrogram-only       & $57.7 \pm 2.1$  & $37.2 \pm 3.1$  & $69.1 \pm 3.9$  & $63.2 \pm 5.1$ \\
Concatenation          & $56.4 \pm 2.4$  & $34.3 \pm 3.7$  & $62.4 \pm 5.2$  & $59.4 \pm 5.3$ \\
\midrule
UAF (mean pooling)     & $59.4 \pm 1.9$  & $\mathbf{39.7 \pm 1.3}$ & $\mathbf{73.1 \pm 6.9}$ & $\mathbf{71.5 \pm 5.7}$ \\
UAF (GRU)              & $58.8 \pm 3.7$  & $37.8 \pm 4.2$  & $68.1 \pm 3.5$  & $61.0 \pm 6.2$ \\
UAF (Transformer)      & $57.0 \pm 3.0$  & $38.4 \pm 4.0$  & $63.3 \pm 11.0$ & $51.9 \pm 22.4$ \\
UAF (Att.\ Pooling)    & $\mathbf{60.3 \pm 2.5}$ & $38.0 \pm 1.3$ & $57.0 \pm 6.2$ & $43.0 \pm 10.4$ \\
\bottomrule
\end{tabular}%
}
\caption{Classification performance under identity-based splits. In SoundWel the pig identity inferred from filename, recording year, and recording team). Macro-F1 is computed over the 13 of 17 classes present in the test partition. In DogBark, dogs named Mac and Zoe are considred as the test set. Metrics are mean~$\pm$~std over 5 independent training runs. Bold shows the best performing model.}
\label{tab:performance}
\end{table*}

\paragraph{Role of temporal architecture}
Leveraging a GRU or Transformer component within our UAF framework does not yield consistent performance gains in comparison with a simple mean pooling (Table~\ref{tab:performance}), and increases run-to-run variance. This indicates that local chunk-level uncertainty estimation is sufficient and that introducing sequence dependencies may cause overfitting rather than generalization. We therefore recommend mean pooling as the default temporal module for uncertainty-aware bioacoustic fusion at current dataset scales and focus our results analysis on this variant in the following sections.

\subsection{Per-Class Classification Analysis}
\label{sec:confusion}

\begin{figure}[t]
    \centering
    \includegraphics[width=0.42\linewidth]{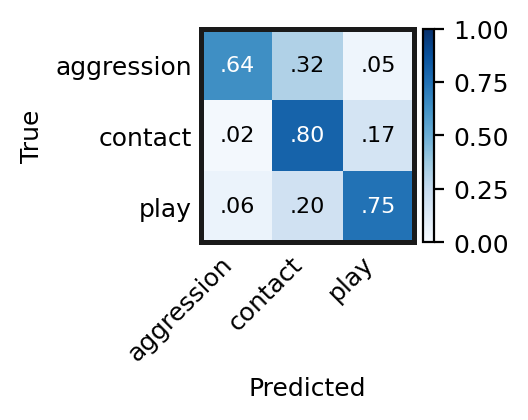}
    \hfill
    \includegraphics[width=0.42\linewidth]{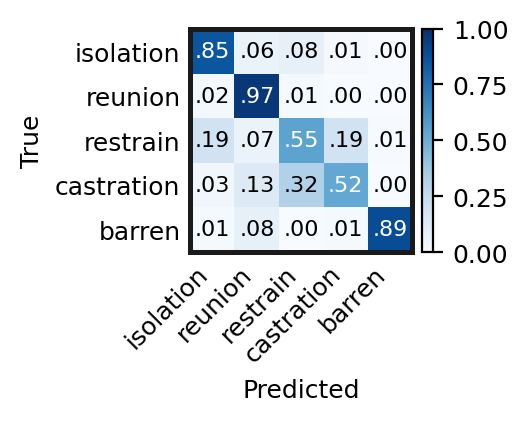}
    \caption{Normalized confusion matrices for UAF (mean pooling) on DogBark (left, 3 classes) and SoundWel (right, 5 welfare-relevant classes). See the full 17-class SoundWel matrix in \ref{sec:supp_confusion_all}, Fig.~\ref{fig:confusion_all_soundwel}.}
    \label{fig:confusion}
\end{figure}

Figure~\ref{fig:confusion} shows normalized confusion matrices for the UAF (mean pooling) model on both datasets. On DogBark (Fig.~\ref{fig:confusion}, left), although there is an imbalance in the number of samples across behavioral classes, recall remains balanced (aggression 0.64, contact 0.80, play 0.75). The largest confusions occur between \textit{aggression} and \textit{contact} barks, both of which are broadband, high-energy temporal calls. For SoundWel (Fig.~\ref{fig:confusion}, right), we show five classes that represent both acute and chronic welfare challenges (see Fig.~\ref{fig:confusion_all_soundwel} for the complete 17-class matrix). \textit{Reunion} is classified near-perfectly. \textit{Restrain} in contrast is the most challenging class, frequently confused with \textit{castration} and \textit{isolation}, which is consistent with the overlapping high-stress acoustic profiles of procedural contexts. The contrast between \textit{barren} (well-recognized) and \textit{restrain} (poorly recognized) is noteworthy: \textit{barren}-housing grunts are tonally distinct, whereas \textit{restrain} calls overlap spectrally with other acute-stress vocalizations. This suggests that the primary remaining challenge on SoundWel lies in distinguishing between welfare-negative calls.

\subsection{Prediction Calibration}
\label{sec:calibration}

\begin{figure}[t]
    \centering
    \includegraphics[width=0.48\linewidth]{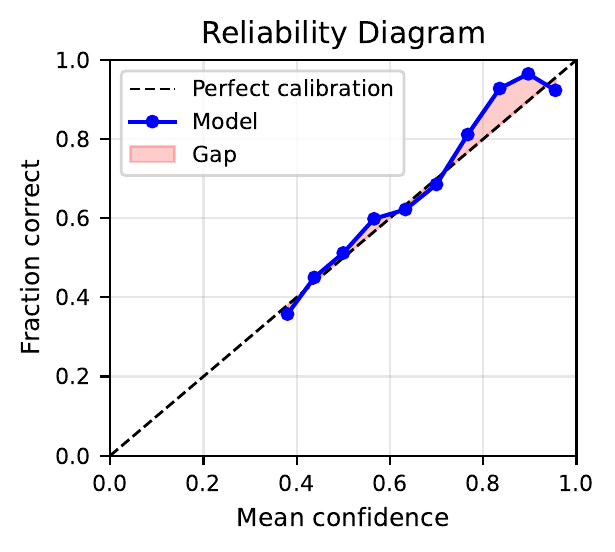}
    \hfill
    \includegraphics[width=0.48\linewidth]{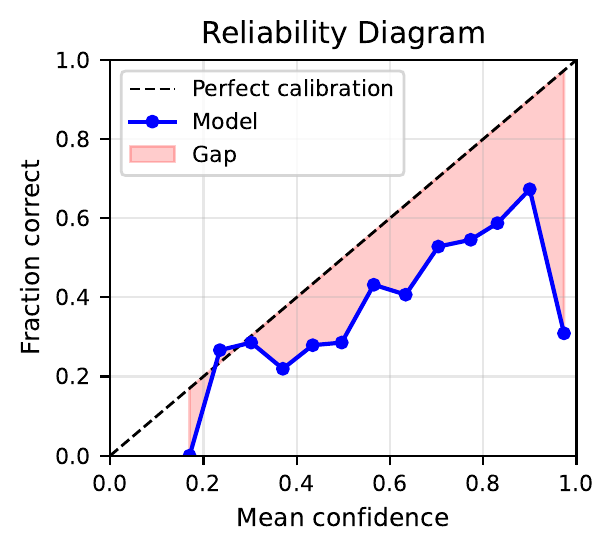}
    \caption{Reliability diagrams for UAF (mean pooling) on DogBark (left) and SoundWel (right). A perfectly calibrated model lies on the dashed diagonal. Points above indicate under-confidence (the model is more accurate than its stated confidence), and points below indicate over-confidence.}
    \label{fig:reliability}
\end{figure}

\begin{figure}[t]
    \centering
    \includegraphics[width=0.65\linewidth]{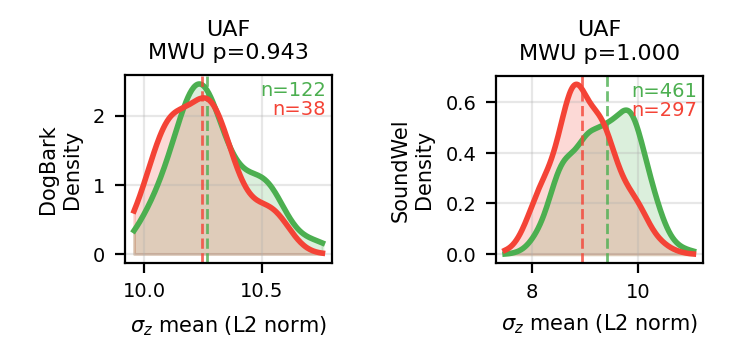}
    \caption{Distribution of $\|\sigma_z\|$ (latent uncertainty norm) for UAF (mean pooling) on DogBark (left) and SoundWel (right). Green: correctly classified recordings; red: incorrectly classified.}
    \label{fig:sigma_dist}
\end{figure}

Beyond improving classification accuracy, the uncertainty-aware design aims to produce confidence estimates that are well calibrated. Hence, a prediction reported with, e.g., 90\% confidence should be correct roughly 90\% of the time. We assess this along several complementary metrics. Figure~\ref{fig:reliability} and ~\ref{fig:sigma_dist} visualize this assessment, while Table~\ref{tab:calibration} quantifies it numerically. The Expected Calibration Error (ECE) measures the average gap between stated confidence and actual accuracy, and ECE$_\text{cal}$ reports this same gap after a simple post-hoc temperature-scaling fix \citep{guo2017calibration}, showing how much of the miscalibration is easily correctable. The AUROC of the latent uncertainty norm $\|\sigma_z\|$ \citep{efron1979bootstrap} tests whether the model's own uncertainty signal can distinguish between correct and incorrect predictions, while the Mann--Whitney U test \citep{mann1947test} checks whether $\sigma_z$ is statistically significantly larger for incorrect predictions than for correct ones. The Brier score \citep{brier1950verification} folds accuracy and calibration into a single proper scoring rule, penalizing confident mistakes most heavily.

The reliability diagrams in Fig.~\ref{fig:reliability} shows that our model is modest about its accuracy for DogBark, while it is overconfident for Soundwel. The reason could be that the DogBark dataset has long calls of the dogs and majority voting over many sliding-window chunks helps to recover the correct label more often than any individual chunk's own confidence would suggest. On the other hand, Soundwel reflects its greater dataset complexity. However, in Table~\ref{tab:calibration}, ECE evaluations show that the reliablity for Soundwel could be corrected at test phase by fitting a scalar temperature. Further, the Brier scores (0.199 for DogBark, 0.217 for SoundWel) indicate the model's performance based on accuracy and uncertainty is well below the worst-case value of 1.

The Difficulty-Confidence loss (see Section \ref{par:DC_loss}) pushes harder samples toward lower values of $\sigma_z$. The AUROC (0.415 for DogBark, 0.337 for SoundWel) and $p_\text{MWU}$ (0.943 for DogBark, 1.000 for SoundWel) values in Table~\ref{tab:calibration} confirm this statement. Consequently, the latent uncertainty could be used as a misclassification signal, such that a low $\sigma_z$ indicates that the recording is most likely misclassified. For example, see Figure \ref{fig:sigma_dist}, where the distribution of correctly classified samples for the SoundWel dataset has a larger mean. However, this is not a strong misclassification signal for the DogBark dataset, which we attribute to the limited size of this dataset, with only 122 correctly classified and 38 misclassified samples.

\begin{table}[htbp]
\centering
\caption{Uncertainty calibration metrics for UAF (mean pooling).}
\label{tab:calibration}
\setlength{\tabcolsep}{4pt}
\begin{tabular}{lccccc}
\toprule
Dataset & ECE$\downarrow$ & ECE$_\text{cal}\downarrow$ & AUROC & Brier$_\text{bin}\downarrow$ & $p_\text{MWU}$ \\
\midrule
DogBark  & 0.195 & 0.109 & 0.415$_{[0.31,0.52]}$ & 0.199 & 0.943 \\
SoundWel & 0.144 & 0.074 & 0.337$_{[0.30,0.38]}$ & 0.217 & 1.000 \\
\bottomrule
\end{tabular}
\end{table}

\subsection{Prediction Confidence by Behavioral Context}
\label{sec:uncertainty}

\begin{figure}[t]
    \centering
    \includegraphics[width=0.38\linewidth]{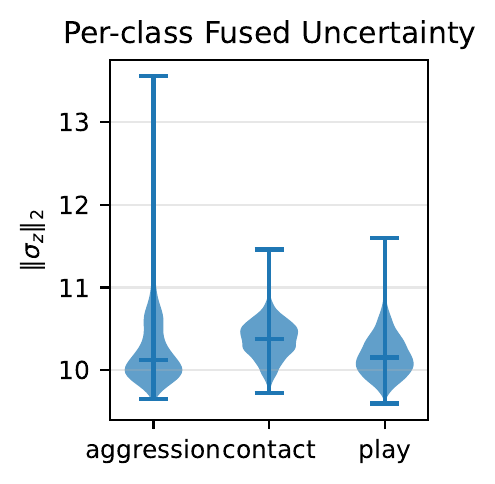}
    \hfill
    \includegraphics[width=0.60\linewidth]{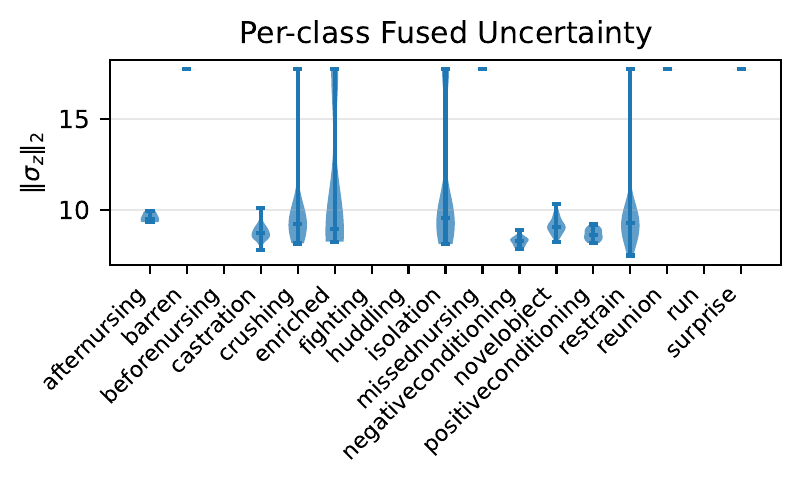}
    \caption{Per-class predicted confidence distributions for UAF (mean pooling) on DogBark (left, 3 classes) and SoundWel (right, 17 classes). Each violin shows the distribution of the model's maximum softmax probability for correctly classified samples of that class.}
    \label{fig:violin}
\end{figure}

Figure~\ref{fig:violin} shows how confident correct predictions are within each behavioral class: a high median confidence with low spread indicates a class that the model recognizes consistently, while a low median confidence or wide spread indicates a class where the  model is unsure between different possibilities, even if it ultimately predicts correctly.

On DogBark, the model produces relatively higher-confidence predictions for aggression barks, which have a distinct broadband impulsive structure, and somewhat lower confidence for contact and play barks. This observation is consistent with the confusion matrix patterns. On SoundWel, confidence distributions are more varied across the 17 classes. These distributions are qualitatively informative for practitioners such that a class with a consistently low-confidence distribution is a candidate for targeted data collection or class-specific fine-tuning.

\subsection{Fusion Weight Adaptation Under Targeted Modality Corruption}
\label{sec:adaptive}

As an additional robustness experiment, we trained a UAF-based model with Gaussian noise injected into one of the streams during training, including complete signal zeroing to simulate missing inputs and noisy barn environments. 

\begin{figure}[htbp]
    \centering
    \includegraphics[width=0.98\linewidth]{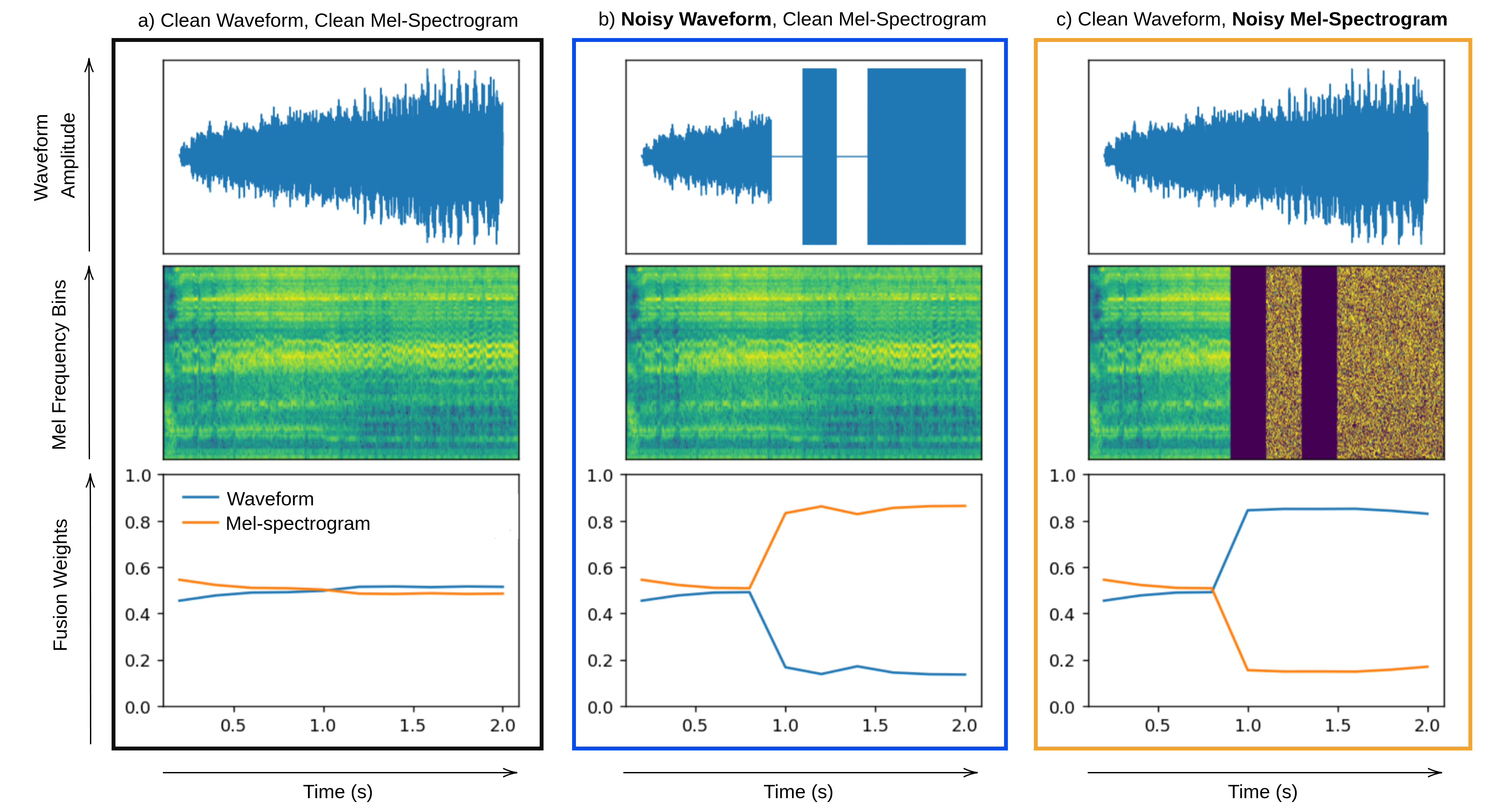}
    \caption{Adaptive modality fusion weights under controlled noise injection for a UAF model trained with deliberative noise augmentation. (a)~Clean waveform and mel-spectrogram: fusion weights remain balanced when both siganls are clean and equally informative. (b)~Clean spectrogram and noisy waveform at 0.6~s: waveform uncertainty rises, shifting fusion weight toward the spectrogram. (c)~Clean waveform and noisy spectrogram at 0.6~s: fusion weight shifts toward the waveform.}
    \label{fig:fusionweights}
\end{figure}

As shown in Fig.~\ref{fig:fusionweights}, in the clean condition (panel~a), fusion weights oscillate near 0.5 for both modalities throughout the recording. When Gaussian noise is injected into the waveform at 0.6~s (panel~b), the variance weighting shifts to the clean spectrogram. The opposite behavior occurs when the spectrogram is corrupted (panel~c) and waveform is the clean input of the model. This indicates that noise injection during training can improve the sensitivity of the fusion weight response in the test phase. 

\section{Conclusion}
\label{sec:conclusion}

We presented Uncertainty-Aware Fusion (UAF), a dual-stream architecture that estimates per-utterance Gaussian uncertainty for waveform and spectrogram representations and fuses them via dimension-wise uncertainty weighting. Under an evaluation on SoundWel and DogBark that excludes all individuals seen during training, UAF (mean pooling) consistently outperforms both single-modality baselines and static-concatenation fusion. A controlled temporal aggregation ablation shows that simple mean pooling mostly outperforms GRU, Transformer, and attention pooling at current dataset scales. This confirms that the performance improvement arises from uncertainty-driven fusion and not the temporal modeling depth. Calibration analysis of UAF (mean pooling) shows that uncertainty gating reliably improves classification accuracy.

While our approach shows promising results, several directions remain for future research. Both datasets contain a limited number of individuals, resulting in a correspondingly small test pool. Therefore, multi-farm benchmarks with broad environmental variability are needed to characterize generalization at deployment scale. Furthermore, we currently focus on two distinct representations of acoustic monitoring, while the proposed method can be extended to additional sensing modalities. Future methodological work could involve collecting a multimodal dataset and extending UAF to multi-farm settings with complementary modalities, such as video or accelerometry, to address limitations in both scalability and multimodal integration.

\section*{Declaration of competing interest}
The authors declare that they have no known competing financial interests or personal relationships that could have appeared to influence the work reported in this paper.

\section*{CRediT authorship contribution statement}
\textbf{Ehsan Yaghoubi:} Conceptualization, Methodology, Software, Validation, Formal analysis, Investigation, Resources, Data curation, Writing -- original draft, Visualization. \textbf{Florian Haselbeck:} Conceptualization, Supervision, Writing -- review \& editing, Funding acquisition.

\section*{Data availability}
Both datasets used in this study are publicly accessible. The full-size unlabeled DogBark dataset \citep{yin2004barking} is archived at \url{https://archive.org/details/dog-barks-raw}. We use the annotated version containing 693 samples, which is available at \url{https://huggingface.co/datasets/cgeorgiaw/animal-sounds}. The SoundWel dataset \citep{briefer2022classification} is provided as a labeled audio database on Zenodo at \url{https://zenodo.org/records/8252482}. All resources were accessed on 27 July 2026.

\section*{Declaration of generative AI and AI-assisted technologies in the manuscript preparation process}
The authors used Google Scholar Labs, Claude Code, ChatGPT, Monica, Grok, Perplexity, and Gemini for improving the clarity of language, brainstorming, literature and dataset search, as well as a critical review of the manuscript. The authors subsequently reviewed and significantly revised all AI-assisted suggestions. The authors assume full responsibility for the content presented in this manuscript.

\section*{Acknowledgments}
We gratefully acknowledge the support of the Free State of Bavaria for funding the positions of EY and FH via the Bavarian Hightech Agenda. The funder had no role in the study design, data collection and analysis, decision to publish, or preparation of the manuscript.

We would like to thank Harshavardan Subramanian and Dr. Georg Back for their constructive discussions and valuable feedback during the development of this work.

\bibliographystyle{elsarticle-harv}   
\bibliography{references}             

\appendix
\setcounter{figure}{0}
\setcounter{table}{0}
\renewcommand{\theHfigure}{\Alph{section}.\arabic{figure}}
\renewcommand{\theHtable}{\Alph{section}.\arabic{table}}

\section{Complete SoundWel Confusion Matrix}
\label{sec:supp_confusion_all}

\begin{figure}[htbp]
    \centering
    \includegraphics[width=0.7\linewidth]{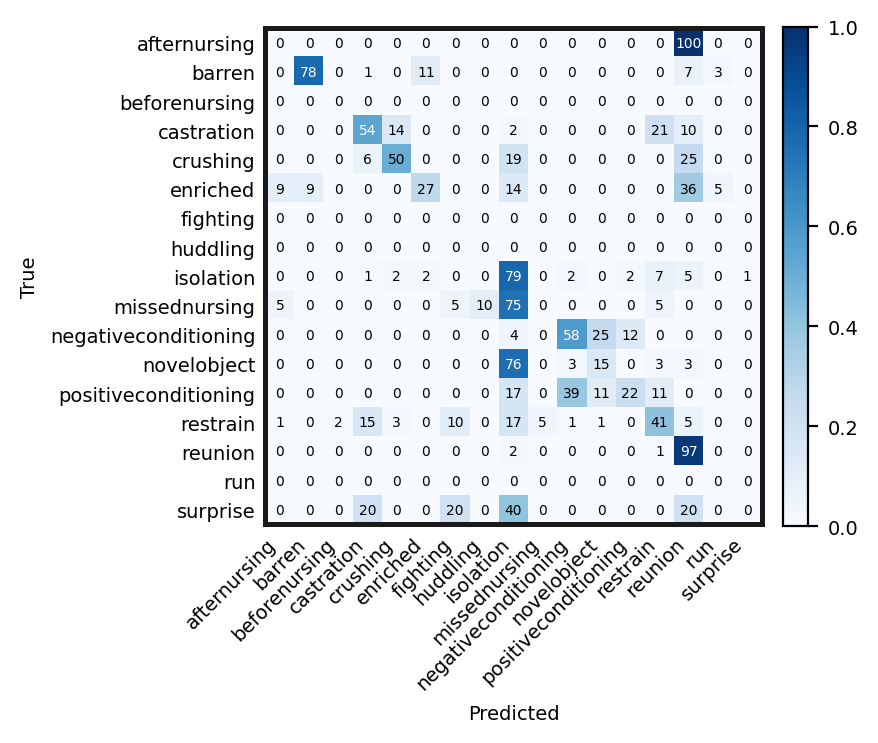}
    \caption{Complete 17-class normalized confusion matrix for UAF (mean pooling) on SoundWel. Cell values show per-class recall as percent. The four classes absent from the test partition (BeforeNursing, Fighting, Huddling, Run) appear as all-zero rows.}
    \label{fig:confusion_all_soundwel}
\end{figure}

Figure~\ref{fig:confusion_all_soundwel} shows the complete 17-class normalized confusion matrix for UAF (mean pooling) on SoundWel, extending the 5-class subset shown in Figure~\ref{fig:confusion}. 

\section{Random Stratified Split Results}
\label{sec:supp}

Table~\ref{tab:sup_random_split} reports SoundWel results under a random stratified 80/20 split, in which no identity constraint is imposed and individual animals may appear in both training and test sets. This protocol is commonly used in the bioacoustics literature and we did it to perform a comparison with prior published results such as \citet{briefer2022classification}.
However, because individual-specific acoustic signatures can leak from training into the test partition, the resulting metrics overestimate the generalization ability of the model to truly unseen individuals.
Under this split, UAF (GRU) achieves 86.0\% accuracy and 76.5\% macro F1, surpassing both the simple static concatenation baseline and \citet{briefer2022classification}.

\begin{table}[htbp]
\centering
\footnotesize
\setlength{\tabcolsep}{8pt}
\begin{tabular}{lcc}
\toprule
\textbf{Model} & Acc.\ (\%) & Macro F1 (\%) \\
\midrule
\citet{briefer2022classification} (reported) & $81.5$ & $81.2$ \\
Concatenation          & $85.5 \pm 0.3$ & $71.6 \pm 1.3$ \\
UAF (GRU)              & $\mathbf{86.0 \pm 0.3}$ & $\mathbf{76.5 \pm 1.1}$ \\
UAF (Transformer)      & $81.8 \pm 0.2$ & $65.5 \pm 1.0$ \\
UAF (Att.\ Pooling)    & $85.7 \pm 0.3$ & $74.7 \pm 1.5$ \\
\bottomrule
\end{tabular}
\caption{SoundWel results under a \emph{random stratified 80/20 split} (no identity separation). Metrics are mean~$\pm$~std over 5 independent runs. Bold: best among the models.}
\label{tab:sup_random_split}
\end{table}

\end{document}